\documentclass[sigconf]{acmart}

\usepackage{enumitem}
\usepackage{subcaption}
\usepackage{multirow}
\usepackage{algorithmic}
\usepackage{amsmath}
\usepackage{dsfont}

\usepackage{xcolor}
\newif\ifrevtracking\revtrackingfalse
\definecolor{cgen}{HTML}{1565C0}    
\definecolor{cdecomp}{HTML}{2E7D32} 
\definecolor{cgpu}{HTML}{E65100}    
\definecolor{cdeploy}{HTML}{6A1B9A} 
\definecolor{cfix}{HTML}{AD1457}    
\newcommand{\rgen}[1]{\ifrevtracking\textcolor{cgen}{#1}\else#1\fi}

\newcommand{\rfix}[1]{\ifrevtracking\textcolor{cfix}{#1}\else#1\fi}

\AtBeginDocument{%
  }

\setcopyright{none} 
\copyrightyear{2026}
\acmYear{2026}

\acmConference[MICRO 2026]{The 58th IEEE/ACM International Symposium on Microarchitecture}{October 31--November 04, 2026}{Athens, Greece}

\acmISBN{978-X-XXXX-XXXX-X/XX/XX}

\acmSubmissionID{2062}

\begin{document}

\title{Sparse by Command: Task-Conditional Compute Skipping for Multi-Task Inference Accelerators}

\author{Afzal Ahmad}
\email{eeafzal@ust.hk}
\affiliation{
  \institution{The Hong Kong University of Science and Technology}
  \country{Hong Kong}
}

\author{Gaoyu Mao}
\email{maogaoyu1@huawei.com}
\affiliation{
  \institution{Huawei Noah's Ark Lab}
  \country{Hong Kong}
}

\author{Shoubo Hu}
\email{hushoubo@huawei.com}
\affiliation{
  \institution{Huawei Noah's Ark Lab}
  \country{Hong Kong}
}

\author{Hui-Ling Zhen}
\email{zhenhuiling2@huawei.com}
\affiliation{
  \institution{Huawei Noah's Ark Lab}
  \country{Hong Kong}
}

\author{Mingxuan Yuan}
\email{yuan.mingxuan@huawei.com}
\affiliation{
  \institution{Huawei Noah's Ark Lab}
  \country{Hong Kong}
}

\author{Xinyu Chen}
\email{xinyuchen@hkust-gz.edu.cn}
\affiliation{
  \institution{The Hong Kong University of Science and Technology (Guangzhou)}
  \city{Guangzhou}
  \country{China}
}

\author{Wei Zhang}
\email{eeweiz@ust.hk}
\affiliation{
  \institution{The Hong Kong University of Science and Technology}
  \country{Hong Kong}
}
    




\begin{abstract}
Multi-task inference models share a single backbone across diverse tasks, yet execute identical computation regardless of which task is active — wasting energy and cycles on task-irrelevant operations. We observe that the task command, typically available before inference begins, provides a free signal that can be exploited to skip unnecessary computation at the hardware level. We present a HW/SW co-designed approach in which a lightweight gating network, trained jointly with the backbone, predicts per-tile binary execution masks conditioned on the task input. Each tile corresponds to a fixed group of output channels (the native scheduling granularity of the accelerator), enabling masked tiles to be skipped with zero overhead. This yields a task-dependent reduction in compute, where each command activates only the subset of the network it requires, without changes to the model architecture or inference pipeline.

We co-design the full system stack: a command-conditioned training procedure that learns hardware-aligned tile masks under a sparsity objective; an instruction set architecture whose instructions carry per-tile bitmask fields, allowing the hardware to skip masked tiles without software intervention; and a tiled inference accelerator with configurable parallelism, double-buffered memory, and INT8 datapath that natively supports sparse tile execution.

We prototype on an AMD/Xilinx Alveo U50 FPGA and evaluate on a closed-loop visuomotor driving task in CARLA autonomous driving simulator. Task-conditional sparsity reduces FLOPs by 66--76\% while maintaining driving quality. On-device latency decreases by 51--59\%, from 9.12 ms to 3.74--4.44 ms (2.1--2.4$\times$ speedup), with energy per inference dropping from \rgen{263 to 108--128\,mJ}.
\end{abstract}



\maketitle

\begin{center}
\small
This is the accepted version of the paper that will appear in the
Proceedings of the 59th IEEE/ACM International Symposium on
Microarchitecture (MICRO 2026).
\end{center}

\section{Introduction} \label{sec:introduction}
Multi-task inference workloads are increasingly common in real-time autonomous systems~\cite{caruana1997multitask, bojarski2016end}, where a single neural network backbone must serve multiple operating modes (lane following, turning, braking in navigation) under strict latency and power constraints~\cite{sze2017efficient}.

These models share the same convolutional or transformer backbone across all tasks, yet the hardware executes identical computation regardless of which task is active. A lane-following command, for instance, may not require the same feature channels as an intersection turn, yet every output channel tile is computed and every weight tile is fetched from memory. This uniform execution wastes both compute cycles and memory bandwidth, resources that are particularly scarce on edge accelerators backed by high-bandwidth memory (HBM).

A natural solution is to exploit sparsity: skip computation that does not contribute to the active task. Prior work on sparse accelerators has made significant progress in this direction. SCNN~\cite{parashar2017scnn} and SparTen~\cite{gondimalla2019sparten} exploit unstructured weight and activation sparsity through compressed encodings and index-matching hardware, while Cambricon-S~\cite{zhou2018cambricon} co-designs coarse-grained pruning with a dedicated index decoder. Eyeriss v2 \cite{chen2019eyeriss} introduces a flexible dataflow that adapts to varying layer shapes. \textbf{However, these designs treat sparsity as a \textit{static} property:} the sparse pattern is determined at training or compile time and does not change based on what the model is being asked to do at runtime.

A parallel line of work on dynamic neural networks has explored input-conditional computation. SkipNet~\cite{wang2018skipnet} and BlockDrop~\cite{wu2018blockdrop} learn per-input gating policies that skip entire residual blocks, while FBS~\cite{gao2019fbs} dynamically suppresses uninformative channels based on intermediate activations. Mixture-of-Experts architectures route tokens to specialized subnetworks. These approaches achieve impressive FLOP reductions, but are designed for GPU or CPU execution where ``skipping'' a block means zeroing its output; the hardware still fetches weights, allocates registers, and occupies memory bandwidth. \textbf{Without structured alignment to the hardware's tiling granularity, dynamic sparsity does not translate to proportional speedup on real accelerators~\cite{mishra2021accelerating}}.

Meanwhile, accelerator ISA design has matured around fixed computation graphs. Gemmini~\cite{genc2021gemmini} exposes matrix-multiply primitives through a RISC-V-integrated instruction set, and NVDLA~\cite{nvdla2017} uses per-layer descriptors to configure convolution, pooling, and activation operations. These ISAs efficiently schedule dense workloads but lack any mechanism to express conditional tile execution; they have no opcode field that tells the hardware which output channel groups to skip based on a runtime signal.

We observe that multi-task models carry a signal that \textbf{none of these prior approaches exploit: the \textit{task command} itself}. In a visuomotor driving controller, the high-level command (follow lane, turn left, brake) is known before inference begins and remains fixed for the duration of that inference. \textbf{This command is a free input that can be used to predict, at \textit{zero runtime cost}, which tiles of the backbone are unnecessary}, enabling structured sparsity that adapts per task without any per-input overhead.

We present a full-stack co-designed system that bridges the gap between dynamic neural networks and sparse hardware accelerators. Our contributions are:

\begin{itemize}[leftmargin=12pt]
\item \textbf{A task-conditional gating mechanism} that predicts per-tile binary execution masks from a task command vector. A lightweight MLP ($<$0.12\% of backbone parameters) is trained jointly with the model through a three-phase pipeline transitioning from soft to hard binary masks.


\item \textbf{An instruction set architecture} with native tile mask support. Each instruction carries per-tile bitmask fields; the hardware skips masked tiles entirely (no weight fetch, no activation load, no compute) at zero control overhead.


\item \textbf{A tiled inference accelerator} with INT8 datapath, configurable parallelism, and double-buffered memory, instrumented with cycle-accurate performance counters. Sparsity is controlled entirely through instruction-level masks with no datapath changes.


\item \textbf{End-to-end closed-loop evaluation} on an AMD/Xilinx Alveo U50 FPGA integrated with the CARLA driving simulator. Task-conditional sparsity reduces FLOPs by 66--76\% across six driving commands while maintaining 100\% route completion.

\end{itemize}

\begin{figure}[t] 
 \centering 
 \includegraphics[width=0.9\columnwidth]{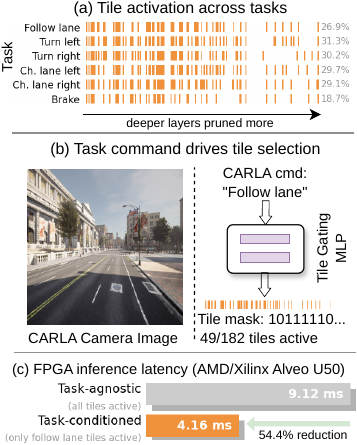} 
 \caption{Task-conditional tile sparsity. (a)~Per-task tile activation masks (orange = active, each column = 16 output channels). (b)~Driving command produces a binary tile mask via a gating MLP. (c)~FPGA inference latency: 54.4\% reduction.} 
 \vspace{-0.2in}
\label{fig:hero} 
\end{figure}

Figure~\ref{fig:hero} illustrates the approach end-to-end: per-task tile activation masks (a) reveal that each driving command activates a distinct subset of the backbone's output channel tiles; a lightweight gating MLP produces these masks from the driving command (b); and on-device FPGA execution with task-conditional masking reduces inference latency by 54.4\% for lane following compared to dense execution (c).
   
To support reproducibility and future research, we release the full co-design stack (RTL, host runtime, training pipeline, and evaluation scripts) as open source\footnote{\url{https://github.com/afzalxo/sparse-by-command}}.

\section{Background and Motivation}

\subsection{Multi-Task Visiomotor Control}
\begin{figure}[t] 
\centering
\includegraphics[width=\columnwidth]{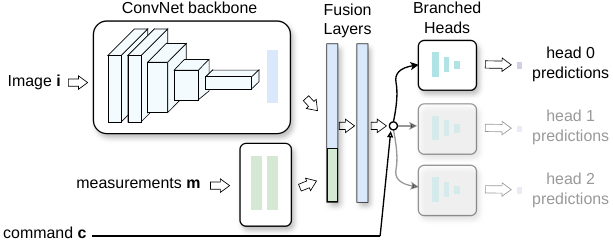}
\vspace{-0.1in}
\caption{Visiomotor Controller: a shared convolutional backbone processes the camera image \textbf{i}, and the driving command \textbf{c} selects one of several branched heads to produce steering and acceleration outputs~\cite{codevilla2018end}. All backbone layers execute fully regardless of the active command.}
\vspace{-0.2in}
\label{fig:model_arch_dense} 
\end{figure}

Visuomotor controllers for autonomous systems map sensor observations directly to continuous control outputs (steering angle, throttle, and braking) conditioned on a high-level navigation command. Unlike modular perception-planning-control pipelines, these end-to-end models~\cite{bojarski2016end, codevilla2018end} use a single neural network (Figure.~\ref{fig:model_arch_dense}) that receives a camera image \textbf{i} and a discrete command \textbf{c} such as "turn left at the next intersection" and regresses the appropriate control action. The command selects one of several parallel output heads, each specialized for a driving behavior, while the backbone is shared across all commands.

This architecture is efficient in parameters but wasteful in compute: the full backbone executes regardless of which command is active. Intuitively, the features required for lane following (road curvature, lane markings) differ from those required for braking (proximity to obstacles, traffic light state). Yet every output feature group in every layer is computed for every inference, even when the active command only requires a subset.

The key insight is that the command is known before inference begins and typically persists for hundreds of consecutive frames. A "turn left" command, for example, remains active throughout the entire turn maneuver. This temporal stability means that a per-command execution mask can be computed once and reused across many inferences, amortizing any gating overhead to near zero.

\subsection{Tiled Execution on Hardware Accelerators}
Modern inference accelerators decompose layer computation into tiles, fixed-size blocks that match the hardware's parallelism dimensions~\cite{chen2016eyeriss, sze2017efficient}. Whether the operation is a convolution or a fully-connected (linear) projection, the accelerator processes the output in groups of output features (output channels in CNNs, columns of the weight matrix in transformers' linear layers), input features, and, for convolutional layers, spatial rows. Each tile execution involves three phases: (1) loading the input activation tile from external memory into an on-chip buffer, (2) loading the corresponding weight tile, and (3) performing the multiply-accumulate computation.

We define a \textbf{\textit{tile}} as one group of \texttt{OC\_PAR} contiguous output channels, spanning all input channels and spatial positions within a horizontal strip. A tile is the atomic unit of scheduling: the accelerator either executes it in full (loading its weights, streaming activations, accumulating results) or skips it entirely. \textbf{Skipping an output feature tile eliminates all three phases for every input feature tile that would have contributed to it:} no weight fetch, no activation load, and no computation. Crucially, this skipping propagates across layers. An output feature group that is masked in layer L does not need to be computed, and consequently does not need to be loaded as an input feature group in layer L+1. The mask on output features thus simultaneously reduces both the output-side and input-side computation of adjacent layers.

This structured approach stands in contrast to fine-grained sparsity based on individual zero weights or activations, which requires index-matching logic and irregular memory access patterns that complicate the datapath~\cite{parashar2017scnn, gondimalla2019sparten}. Tile-level sparsity aligns with the accelerator's parallelism boundaries by construction, enabling simple control flow through a bitmask check rather than complex sparse encoding hardware.

While we describe and evaluate our approach using convolutional layers, the same tiling strategy applies directly to the linear projections in transformer architectures, where the output and input embedding dimensions play the same role as output and input feature groups. The tile mask mechanism is agnostic to the operation type: it controls which groups of output features are computed, regardless of whether the underlying operation is a convolution or matrix multiplication.

Existing tiled accelerators, however, execute all tiles unconditionally. The tile iteration loop is controlled by the layer dimensions encoded in the instruction stream, with no mechanism to selectively skip tiles based on runtime information.

\begin{figure}[t] 
\centering
\includegraphics[width=\columnwidth]{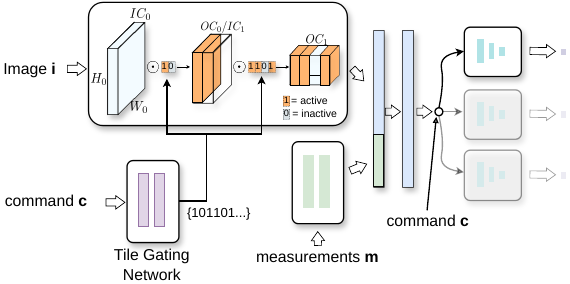}
\vspace{-0.1in}
\caption{Task-conditional model: the command is additionally passed through a tile gating network that produces a binary mask over the backbone's output feature tiles. Masked tiles (inactive) are skipped during execution, reducing both compute and memory traffic.} 
\vspace{-0.2in}
\label{fig:model_arch_gated} 
\end{figure}

\begin{figure*}[t]
\centering
\includegraphics[width=0.80\textwidth]{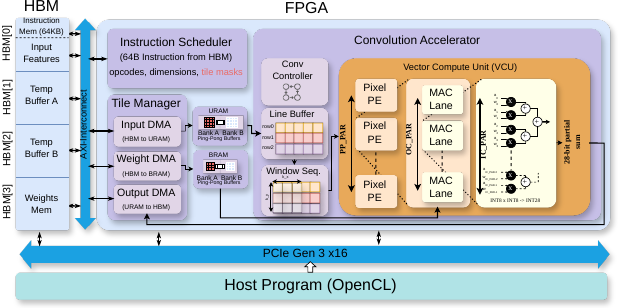}
\vspace{-0.1in}
\caption{Accelerator architecture. The instruction scheduler fetches 64-byte instructions (including tile masks) from HBM. The tile manager orchestrates DMA engines and the compute pipeline. Activations and weights are double-buffered. } 
\vspace{-0.15in}
\label{fig:accelerator}
\end{figure*}

\subsection{The Opportunity: Command-Conditional Tile Masks}
We identify a gap at the intersection of dynamic neural networks and tiled hardware accelerators. Dynamic networks (Section~\ref{sec:introduction}) demonstrate that computation can be safely skipped based on runtime signals, but they operate at granularities (individual weights, attention heads, residual blocks) that do not map to hardware tile boundaries. Tiled accelerators provide the execution mechanism for structured skipping, but \textit{lack the runtime signal} to decide what to skip.

The task command bridges this gap. It is available before inference begins, imposing \textit{no runtime overhead}. It is discrete and \textit{low-dimensional}: a one-hot vector over a small set of commands (six in our driving application). It is \textit{temporally stable}, persisting across many consecutive inferences. And it is \textit{semantically meaningful}, as different commands genuinely require different features. 

 While we demonstrate this approach primarily on a CNN-based driving controller, the mechanism extends to command-conditioned multi-task models with a discrete task selector. \rgen{We confirm this on a ViT-Base backbone (Section~\ref{sec:generality_vit}): the same gating and bitmask mechanism masks the MLP linear-layer tiles of a transformer, and masking attention head-groups is a complementary extension.} In robotic manipulation, grasp-type commands (pinch, scoop, push) could serve as the task signal. The key requirements are: (1) a discrete, low-dimensional task descriptor known before inference, and (2) an accelerator that schedules computation in fixed-size output channel groups. A lightweight gating network can map the command vector to a binary tile mask in a single forward pass through a small MLP (Figure.~\ref{fig:hero} (b), Figure.~\ref{fig:model_arch_gated}), a computation that takes nanoseconds on any processor and produces a mask that remains valid for the duration of the command. The mask is then encoded into the instruction stream as a bitmask field, requiring no changes to the accelerator datapath. 
 
For a backbone with $L$ layers and $T_l$ output feature tiles per layer, the gating network produces a binary vector of length sum($T_l - 1$), one bit per tile, excluding tile 0 of each layer which is always active. This first tile serves as a shared "polysemantic core" that preserves cross-task representations. The total number of predictable tiles in our 8-layer convnet backbone is 182, and the gating MLP has fewer than 13k parameters, less than 0.12\% of the backbone.

\section{System Design}
Our system co-designs three layers: a \textit{software gating mechanism} that produces per-tile binary masks from the task command, an \textit{instruction set architecture} that encodes these masks into the hardware instruction stream, and a \textit{tiled inference accelerator} that executes or skips tiles based on the mask bits. We describe each component in turn.

\subsection{Task-Conditional Gating Network} \label{sec:task-cond-gating}
The gating network is a lightweight MLP that maps a task descriptor to a binary execution mask over the accelerator's tile space (Figure.~\ref{fig:model_arch_gated}). The command \textbf{c} input is a one-hot vector over the set of supported commands (six in our driving application: follow lane, turn left, turn right, change lane left, change lane right, and brake). The output is a sigmoid-activated vector of length 182, corresponding to one predictable tile per output feature group across the backbone's eight layers.

\textbf{Tile structure.} The accelerator groups output features into fixed-size tiles of 16 features each (\texttt{OC\_PAR}). An 8-layer backbone with output feature counts [32, 64, 128, 256, 512, 512, 512, 1024] yields [2, 4, 8, 16, 32, 32, 32, 64] tiles per layer, totaling 190 tiles. Tile 0 of each layer is designated as always-on, serving as a shared representational core across all tasks. The remaining 182 tiles are predictable by the gating network.

\textbf{Training pipeline.} We train the model in three phases:

\emph{Phase 1 (Dense).} The backbone and output heads are trained from scratch without the gating network (Figure.~\ref{fig:model_arch_dense}), using standard supervised learning on the driving dataset. This produces a fully capable dense model. We train for 200 epochs.

\emph{Phase 2 (Soft masks).} The gating network is attached and trained jointly with the backbone (Figure.~\ref{fig:model_arch_gated}). The Phase~2 objective augments the task loss with a sparsity penalty on the continuous mask activations:

\begin{equation}
L_{\text{soft}} = L_{\text{task}}(\hat{y}, y) + \lambda \cdot \frac{1}{T}\sum_{i=1}^{T} \sigma(g_i)
\end{equation}

where $g_i$ are the gating network's logits for each of the $T=182$ predictable tiles, $\sigma$ is the sigmoid function, and $\lambda$ ramps linearly from 0 to 0.002 over the 50 finetuning epochs. This sparsity penalty is analogous to group regularization methods for structured pruning~\cite{wen2016learning}, applied here at tile granularity. The masks remain continuous, allowing gradient flow through the gating network. The backbone is initialized from the Phase~1 checkpoint; the gating parameters are initialized randomly. This phase learns which tiles are dispensable for each task while the backbone adapts to operate with reduced capacity.

\emph{Phase 3 (Hard masks).} The masks are binarized via a hard threshold:

\begin{equation}
m_i = \mathds{1}[\sigma(g_i) > 0.5]
\end{equation}
and the Phase~3 objective becomes:
\begin{equation}
L_{\text{hard}} = L_{\text{task}}(\hat{y}_{\text{masked}}, y) + \lambda \cdot \frac{1}{T}\sum_{i=1}^{T} m_i
\end{equation}

where $\hat{y}_{\text{masked}}$ is the model output with binary masks applied. Since $m_i$ is non-differentiable, we use the straight-through estimator~\cite{bengio2013ste} to approximate the gradient: $\partial m_i / \partial g_i \approx \partial \sigma(g_i) / \partial g_i$. Training continues for 50 epochs, hardening the mask decisions and fine-tuning the backbone to perform well under discrete tile execution.

\textbf{Sparsity control.} The hyperparameter $\lambda$ controls the sparsity level. Higher values push more tiles toward zero, increasing the compute reduction at the cost of task accuracy. In our experiments, $\lambda = 0.002$ produces an average of 28\% active tiles ($\sim$72\% pruned) across tasks while maintaining driving fidelity in CARLA. The sparsity is not uniform across tasks: the gating network learns that left turns require more representational capacity (28.9\% of FLOPs active) than braking (23.9\%), reflecting the inherent complexity of each maneuver in right-hand traffic. 

\subsection{Accelerator Architecture} \label{sec:accel-arch}
The accelerator is a tiled inference engine (Figure~\ref{fig:accelerator}) that processes layers by iterating over output feature tiles, input feature tiles, and spatial tile strips. It connects to HBM through three AXI4 master ports: one for reading input activations, one for reading weights and biases, and one for writing output activations. All data uses INT8 representation with 28-bit internal accumulators to preserve precision across partial sum accumulation. The accelerator is parameterized by three parallelism factors: \texttt{OC\_PAR} (output channels per tile, determining tile width), \texttt{IC\_PAR} (input channels processed per PE in one MAC operation), and \texttt{PP\_PAR} (pixel positions processed in parallel across the Vector Compute Unit (VCU) array). Our implementation uses \texttt{OC\_PAR}$=$\texttt{IC\_PAR}$=$16 and \texttt{PP\_PAR}$=$8, yielding 2,048 INT8 MACs per cycle, albeit these parameters are configurable.

\textbf{Tile Manager.} The tile manager is the central orchestrator. For each layer, it iterates over the spatial dimension in strips of configurable height (\texttt{H\_TILE}$=$8 rows in our implementation), then over output feature tiles, then over input feature tiles. For each (output tile, input tile) pair, it coordinates three operations: loading the input activation strip from HBM into an on-chip buffer, loading the corresponding weight tile, and dispatching the data to the compute pipeline. The tile manager implements a double-buffered pipeline: while the compute engine processes one tile from buffer bank A, the DMA engines load the next tile into bank B. When both complete, the banks swap and the cycle repeats. Figure.~\ref{fig:tile_skip_pseudo} shows a psuedocode of the tile manager's execution loops. 

\textbf{DMA Engines.} Four DMA engines move data between HBM and on-chip memory: an input DMA loads activation strips into URAM (burst AXI read, up to 8 outstanding transactions, with zero-padding for halo rows), a weight DMA and bias DMA share a second AXI port to load kernel tiles into BRAM and bias vectors into registers, and an output DMA reads accumulated results, applies quantization (bias addition, ReLU, right-shift, INT8 clamping), and writes the result to HBM. The bias and input DMA operate in parallel on separate AXI ports, hiding bias load latency behind the input transfer.


\textbf{On-Chip Buffers.} The accelerator uses four on-chip storage structures, all double-buffered to support pipeline overlap: activation buffers in URAM (one spatial strip of tile height $+$ 2 halo rows (padding) per bank), weight buffers in BRAM (one 3$\times$3 kernel tile per bank), bias registers (one \texttt{OC\_PAR}-wide vector), and result accumulators (28-bit partial sums across one spatial strip). Partial sums accumulate across input feature tiles; the result accumulator adds to the existing value when processing subsequent input tiles for the same output tile.



\textbf{Compute Pipeline.} The convolution accelerator processes data through a multi-stage pipeline. A line buffer stores the most recent 3 rows of the activation strip, feeding a window sequencer that extracts overlapping 3x3 windows as the data streams through. Each window is broadcast to an array of VCUs, one per pixel position in the parallelism group. Each VCU contains an array of processing elements (PEs), one per output feature in the tile. Each PE performs a multiply-accumulate operation across all input features using a dedicated MAC lane: it multiplies the input feature vector element-wise with the corresponding weight vector and reduces the products through an adder tree to produce a single 28-bit partial sum per cycle.

The total parallelism is \texttt{PP\_PAR}\ $\times$\ \texttt{IC\_PAR}\ $\times$\ \texttt{OC\_PAR} MAC operations per cycle. In our configuration (8 pixels, 16 output features, 16 input features), this yields 2,048 INT8 MAC operations per cycle.

\textbf{Performance Counters.} The accelerator is instrumented with cycle-accurate performance counters that decompose execution time into compute cycles (convolution pipeline active), memory cycles (any AXI transfer active), overlap cycles (both compute and memory active simultaneously, indicating effective double buffering), and stall cycles (neither active, indicating pipeline bubbles). The stall cycles are further broken down by tile manager state: prologue stalls (initial tile loading), barrier stalls (compute waiting for DMA or vice versa), and output stalls (waiting for the output write to complete). These counters are written to a reserved HBM location at the end of execution and read by the host for analysis. 

\subsection{Neural Instruction Set Architecture} \label{sec:nisa}
The host communicates with the accelerator through a stream of fixed-width instructions stored in HBM. Each instruction is 64 bytes (512 bits), aligned to the HBM word width, and fully describes the execution of one layer. The instruction scheduler fetches instructions sequentially from a reserved HBM region and configures the tile manager for each layer.

\textbf{Instruction format.} Each instruction encodes the layer's operation type, dimensions, quantization parameters, memory offsets, and tile execution masks (Table.~\ref{tab:nisa_format}):

\begin{table}[t]
\centering\small
\setlength{\tabcolsep}{3pt}
\caption{NISA instruction format (64 bytes = 512 bits).}
\vspace{-0.1in}
\label{tab:nisa_format}
\begin{tabular}{lrl}
\toprule
\textbf{Field} & \textbf{Bits} & \textbf{Description} \\
\midrule
Memory offsets     & 256 & input, output, weight, bias (64b each) \\
Spatial dims       & 32  & width, height (16b each) \\
Feature dims       & 32  & in\_channels, out\_channels (16b each) \\
Opcode             & 8   & \texttt{CONV}, \texttt{GEMM}, \texttt{MEMCPY}, \texttt{GAP}, \texttt{HALT} \\
Quant shift        & 8   & Post-accumulation right-shift \\
Bank select        & 8   & Ping-pong buffer (Heap/BufA/BufB) \\
Stride             & 8   & Convolution stride \\
Log$_2$ tile height & 8  & Spatial tile height parameter \\
Aux flags          & 8   & ReLU, flatten, \texttt{\color{teal}sparse\_en}, \texttt{\color{black}bias\_en} \\
\emph{Reserved}    & \emph{16} & \emph{Alignment padding} \\
\color{teal}Tile masks $M_{oc}$ & \color{teal}128 & \color{teal}Two 64-bit OC tile bitmasks (lo/hi) \\
\bottomrule
\end{tabular}
\vspace{-0.15in}
\end{table}

\begin{figure}[t]
\begin{algorithmic}[1]
\small
\REQUIRE Layer config $(H, W, IC, OC, stride, shift)$
\STATE $N_{oc} \leftarrow \lceil OC / \text{OC\_PAR} \rceil$  \COMMENT{number of output tiles}
\STATE $N_{ic} \leftarrow \lceil IC / \text{IC\_PAR} \rceil$  \COMMENT{number of input tiles}
\REQUIRE {\color{teal}Output tile mask $M_{oc}[0..N_{oc}{-}1]$ from NISA instruction}
\REQUIRE {\color{teal}Input tile mask $M_{ic}[0..N_{ic}{-}1]$ $\leftarrow$ previous layer's $M_{oc}$}
\FOR{each spatial strip $y$ in $\lceil H / \text{TILE\_H} \rceil$}
  \STATE DMA: load input activation strip (HBM $\rightarrow$ URAM)
  \FOR{each output tile $t_{oc} = 0$ \TO $N_{oc}-1$}
    \IF{{\color{teal}\texttt{sparse\_en} \&\& $M_{oc}[t_{oc}] = 0$}}
      \STATE {\color{teal}\textbf{skip}: no DMA, no compute, 1-cycle advance}
    \ELSE
      \STATE DMA: load bias vector (HBM $\rightarrow$ register file)
      \FOR{each input tile $t_{ic} = 0$ \TO $N_{ic}-1$}
        \IF{{\color{teal}\texttt{sparse\_en} \&\& $M_{ic}[t_{ic}] = 0$}}
          \STATE {\color{teal}\textbf{skip}: input masked from prev layer}
        \ELSE
          \STATE DMA: load weight tile (HBM $\rightarrow$ BRAM)
          \STATE Stream activation window through conv pipeline
          \IF{$t_{ic} = $ first active}
            \STATE Initialize partial sums in result buffer
          \ELSE
            \STATE Accumulate into existing partial sums
          \ENDIF
        \ENDIF
      \ENDFOR
      \STATE Add bias, right-shift by $shift$, clamp to INT8, ReLU
      \STATE DMA: write output tile (URAM $\rightarrow$ HBM)
    \ENDIF
  \ENDFOR
\ENDFOR
\end{algorithmic}
\vspace{-0.1in}
\caption{Tile manager execution loops. {\color{teal}Teal} highlights the sparsity mechanism. $M_{oc}$ is from the NISA bitmask field; $M_{ic}$ is set to the previous layer's $M_{oc}$ by the instruction scheduler, propagating sparsity across layers. Masked tiles cost one cycle (shift register advance); this two-level skipping compounds across adjacent layers.
}
\vspace{-0.2in}
\label{fig:tile_skip_pseudo}
\end{figure}

\textbf{Tile mask encoding.} The two 64-bit mask fields provide 128 mask bits, supporting up to 128 output feature tiles per layer. Bit k corresponds to output feature tile k: if the bit is 1, the tile manager executes that tile; if 0, it skips the tile entirely. For dense execution, all bits are set. For task-conditional sparse execution, the host evaluates the gating network for the active command and writes the resulting binary mask into the instruction fields before launching the accelerator.

This design requires no runtime mask computation on the accelerator. \rgen{As the command set is bounded (six here), we precompute the masks on the host and the gating MLP never runs at inference; the deployed table is a 128-byte mask per command (a 128-bit lo/hi field per backbone layer). On a command change the host rewrites the mask field in each layer's 512-bit instruction, a 512\,byte host-to-HBM transfer measured as 13.4\,$\mu$s on the U50 against a 3.74--4.44\,ms inference---under a percent of one frame, and amortized to near zero over the hundreds of frames a command persists. An unbounded command set (e.g.\ language-conditioned VLAs) would run the MLP once per change, but its forward pass is negligible against a backbone inference.}



\textbf{Instruction stream generation.} The host runtime parses a model specification that lists each layer's dimensions, quantization parameters, weight and bias file locations, and tile masks. It packs one instruction per layer into a contiguous buffer and writes the buffer to HBM. Execution begins when the host writes a start signal to an AXI4-Lite control register, and completes when the accelerator writes performance counters to a reserved HBM location. 

\textbf{Generality.} The ISA is not specific to any model architecture. The same instruction format handles convolutions, fully-connected layers, pooling, and data movement through the opcode field. Adding a new operation type requires defining a new opcode and implementing the corresponding datapath in the tile manager, without changes to the instruction format or mask encoding. The tile mask fields are opcode-independent and apply uniformly to any operation that iterates over output feature tiles.

\subsection{Tile Skipping}
The tile manager's iteration loop checks the mask before processing each output feature tile (Figure.~\ref{fig:tile_skip_pseudo}, line 6). When the sparse execution flag is set in the instruction's auxiliary fields (Table.~\ref{tab:nisa_format} \texttt{sparse\_en}), the tile manager maintains a shift register initialized with the tile mask $M_{oc}$. At each output tile iteration, it inspects the least-significant bit: if the bit is 1, the tile proceeds through the normal execution pipeline (DMA load, compute, accumulate). If the bit is 0, the tile manager advances to the next output tile without issuing any DMA requests or starting the compute engine. The shift register advances by one bit per iteration, requiring no address decoding or lookup table.

\textbf{Propagation across layers.} When an output feature tile is masked in layer L, the corresponding features are never written to HBM. In layer L+1, these features appear as input feature tiles. The tile manager for layer L+1 checks the input feature mask $M_{ic}$ (derived from layer L's output mask) and skips the corresponding input tile iterations (Figure.~\ref{fig:tile_skip_pseudo}, lines 11-12). This means that a single masked output tile in layer L eliminates both the output-side computation in layer L and the input-side computation in layer L+1, compounding the savings.

\textbf{Zero overhead.} Skipping a tile costs one cycle for the mask bit check and shift register advance. No DMA transactions are initiated, no compute resources are occupied, and no on-chip buffer space is consumed. The skipped tile is invisible to the rest of the pipeline. The double-buffered execution continues seamlessly: after skipping one or more masked tiles, the tile manager proceeds to the next active tile and resumes the normal DMA-compute overlap.

\textbf{Interaction with double buffering.} The tile manager searches ahead for the next active tile before entering the pipeline stage. If the next tile is masked, the search advances past it in a single cycle per skipped tile. Once an active tile is found, the pipeline resumes with the standard bank-swap protocol. This lookahead ensures that the double-buffered pipeline does not stall due to skipped tiles; the DMA for the next active tile begins as soon as the current active tile's compute starts.

\subsection{Deployment Pipeline}\label{sec:deployment-pipe}

Deploying a trained PyTorch model on the accelerator requires bridging the gap between floating-point software representations and the accelerator's INT8 tiled datapath. We implement this as a four-stage offline compiler.

\textbf{Quantization.} A calibration pass runs 300 training samples through the model, recording the maximum absolute activation at each layer output. Per-layer quantization shifts are computed as $shift = \lceil \log_2(act\_max \times cumulative\_scale \,/\, Q_{\max}) \rceil$, where $Q_{\max} = 127$ and the cumulative scale tracks the product of input and weight scales through the network. Weights are quantized to INT8 via per-tensor symmetric scaling~\cite{jacob2018quantization}; biases are quantized to INT32 and clamped to the 28-bit accumulator range.

\textbf{Data packing and memory layout.} Feature maps are repacked from PyTorch's planar [C, H, W] to a tiled layout [H\_tiles, C\_slots, H\_local, W, C\_local] where each innermost group of \texttt{IC\_PAR}$=$16 channels aligns with the PE array. Weights are similarly repacked to [OC\_tiles, IC\_tiles, K$_y$, K$_x$, OC\_par, IC\_par]. The compiler lays out all data in a contiguous HBM region: 64\,KB for instructions, followed by the packed input image, then all layers' weights and biases. Per-layer byte offsets are recorded in each NISA instruction.

\textbf{Instruction generation.} For each layer, the compiler emits one 64-byte NISA instruction, populating all fields described in Section~\ref{sec:nisa}. Bank selection alternates the output buffer between BufA and BufB across layers (ping-pong). For task-conditional models, the per-layer binary masks produced by the gating network are encoded into the instruction's tile mask fields. A final HALT instruction terminates the stream. The host program is written in C++ using OpenCL 1.2 bindings linked against AMD/Xilinx's XRT runtime. It transfers the packed input image, weights, biases, and instruction stream to HBM over PCIe Gen3 x16, issues a start signal via an AXI4-Lite control register write, and polls for completion. After execution, the host reads back the output activations and hardware performance counters from reserved HBM locations.

\section{Evaluation}

\begin{figure}[t]
\centering
\includegraphics[width=0.95\columnwidth]{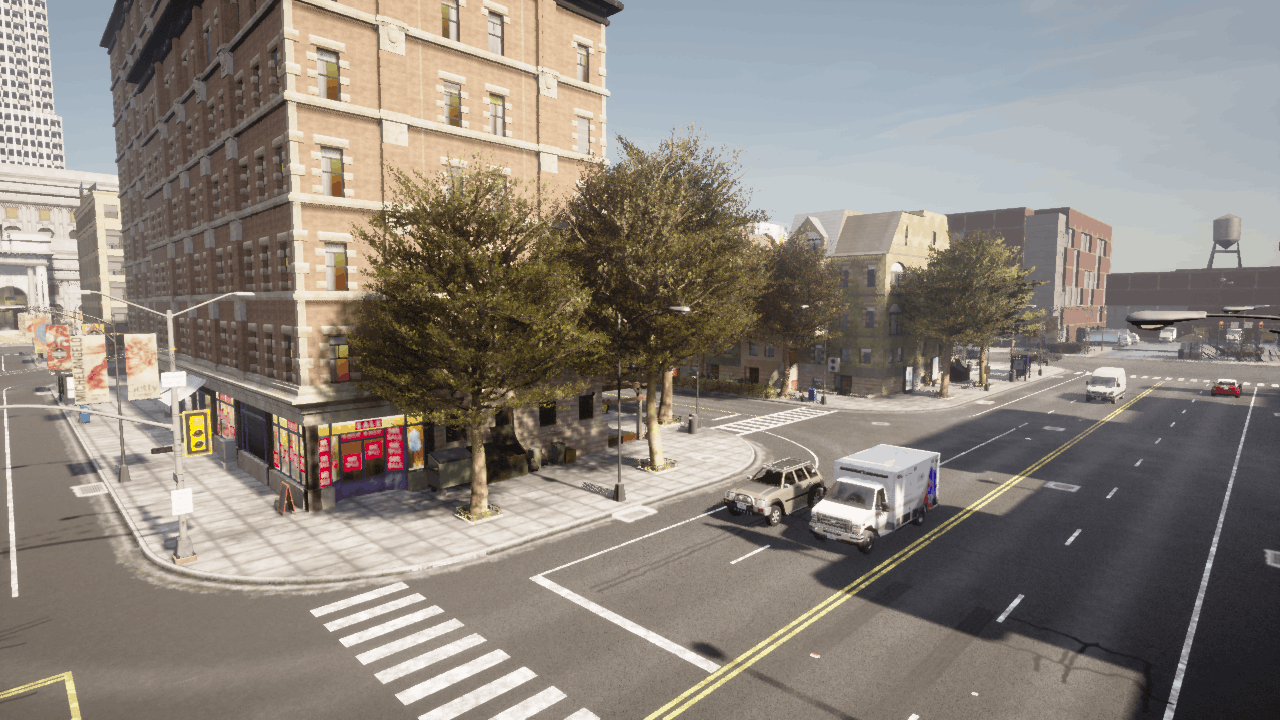}
\vspace{-0.1in}
\caption{Street-level view of an intersection in CARLA \texttt{Town10HD\_Opt}. Multi-lane roads, traffic signals, and urban geometry representative of real-world driving conditions.}
\vspace{-0.2in}
\label{fig:eval_setup_street}
\end{figure}

\subsection{Experimental Setup}

\textbf{Driving task and dataset.}
We evaluate on a closed-loop visuomotor driving task in the CARLA autonomous driving simulator~\cite{dosovitskiy2017carla}, specifically in \texttt{Town10HD\_Opt}, an urban environment with intersections, multi-lane roads, and traffic signals (Fig.~\ref{fig:eval_setup_street}). For dataset collection, an expert rule-based controller drives the ego vehicle at 15\,km/h while a front-facing RGB camera (110$^\circ$ FOV, rendered at 1920$\times$1080) records frames at 20\,FPS, which are subsequently downsampled to 256$\times$256 for training. Each frame is paired with the expert's control labels: steering $\in[-1,+1]$ (where $-1$ corresponds to full left lock and $+1$ to full right lock), throttle $\in[0,1]$, brake $\in[0,1]$, and the vehicle's current speed. To address covariate shift in imitation learning, data collection includes DAgger-style perturbations~\cite{ross2011dagger}: the vehicle is periodically steered off-centre ($\pm$0.20 steering for 15--25 frames) and only the subsequent recovery frames are recorded, teaching corrective behaviour from off-centre states. Since the vast majority of raw frames correspond to the vehicle driving straight, leading to a severely imbalanced command distribution that causes the model to collapse onto a ``drive straight'' policy. To correct this, we drop 90\% of straight-driving frames during dataset curation. The resulting dataset contains 302K training frames across six commands: follow lane / go straight, turn left, turn right, change lane left, change lane right, and brake.

\textbf{Metrics.}
We evaluate closed-loop driving using two metrics. \emph{Route completion} (\%) measures the fraction of planned route waypoints reached before episode termination. \emph{Cross-track error} (CTE, metres) measures the average lateral distance from the vehicle centre to the nearest route waypoint across all simulation frames of an episode. Episodes terminate on collision, sidewalk departure, CTE exceeding 20m, or vehicle standstill for more than 5s without a traffic hazard.

\begin{table*}[t]
\centering
\caption{Per-task sparsity profile and FPGA on-device performance at 287\,MHz. Active Tiles: number of the 182 prunable conv tiles executed for this command (out of 190 total including 8 always-on tiles). FLOP/Param Act.: fraction of dense FLOPs/parameters actually computed. Cycle columns in units of $10^6$ cycles (M); latency derived at 287\,MHz.}
\vspace{-0.1in}
\label{tab:sparsity_latency}
\resizebox{\textwidth}{!}{%
\begin{tabular}{lcccccccc}
\toprule
\textbf{Command} & \textbf{Active Tiles} & \textbf{FLOP Act.} &
\textbf{Param Act.} &
\textbf{Total Cyc.\ (M)} & \textbf{Compute (M)} & \textbf{Stall (M)} &
\textbf{Lat.\ (ms)} \\
\midrule
Dense (all tiles)    & 182/182 (100\%) & 100.0\% & 100.0\% & 2.62 & 1.65 & 0.77 & 9.12 \\
\midrule
Follow Lane / Straight & 49/182 (26.9\%) & 30.2\% & 26.9\% & 1.19 & 0.54 & 0.58 & 4.16 \\
Turn Left            & 57/182 (31.3\%) & 28.9\% & 31.3\% & 1.20 & 0.53 & 0.58 & 4.19 \\
Turn Right           & 55/182 (30.2\%) & 33.7\% & 30.2\% & 1.27 & 0.61 & 0.58 & 4.44 \\
Ch.\ Lane Left       & 54/182 (29.7\%) & 29.9\% & 29.7\% & 1.20 & 0.54 & 0.58 & 4.19 \\
Ch.\ Lane Right      & 53/182 (29.1\%) & 27.6\% & 29.1\% & 1.17 & 0.51 & 0.58 & 4.08 \\
Brake                & 34/182 (18.7\%) & 23.9\% & 18.7\% & 1.07 & 0.43 & 0.57 & 3.74 \\
\bottomrule
\end{tabular}
\vspace{-0.25in}
}
\end{table*}

\textbf{Model.}
The backbone (Figure.~\ref{fig:model_arch_dense}) is an 8-layer stride-2 CNN reducing 256$\times$256$\times$3 input to a 1024-dim feature vector. A measurement encoder projects the previous control state (steering, acceleration, speed) to a 128-dim embedding fused with the visual embedding (1152$\rightarrow$512, ReLU, 50\% dropout). Six parallel branch heads (512$\rightarrow$256$\rightarrow$256$\rightarrow$2, Tanh) output steering and acceleration, each $\in[-1,+1]$ (acceleration$<$0 denotes braking, acceleration$>$0 denotes throttle); the active command selects the branch at runtime. The tile gating MLP (6$\rightarrow$64$\rightarrow$182, sigmoid) adds $<$13k parameters. Total: 11.01\,M parameters and 1.07\,GFLOPs per inference.

\textbf{Training.}
We follow the three-phase pipeline described in Section~\ref{sec:task-cond-gating}. \textbf{Phase~1} trains the dense model (Figure.~\ref{fig:model_arch_dense}) for 200 epochs (AdamW, lr$=$2$\times$10$^{-4}$, BS$=$256) with a branched L1 loss in which, for each sample, only the branch head corresponding to the ground-truth command contributes to the gradient. The steering component is weighted 25$\times$ more heavily than acceleration and is further multiplied by a quadratic magnitude factor $(1+10\,|s|^2)$, where $s$ is the ground-truth steering target. This factor is essential because the dataset is dominated by low-steering frames (straight segments even after the 90\% downsampling); without it, the model minimises loss by predicting near-zero steering uniformly and fails to execute turns. The quadratic weighting penalises large-steering errors much more than straight-driving errors, forcing the model to fit the long tail of the steering distribution. Active braking/throttle frames are additionally up-weighted 5$\times$ relative to coasting frames.  \textbf{Phase~2} attaches the gating network (Figure.~\ref{fig:model_arch_gated}) and fine-tunes for 50 epochs with soft masks under a sparsity penalty ($\lambda{=}$0.002, linearly ramped), reducing active tiles from 93\% to $\sim$80\% of the 182 prunable tiles. \textbf{Phase~3} binarises masks at threshold 0.5 using the straight-through estimator~\cite{bengio2013ste} and fine-tunes for a further 50 epochs. The best Phase~3 checkpoint achieves 100\% route completion and 0.950m average CTE on dynamic hard evaluation routes, with $\sim$72\% of tiles masked across the eight convolutional layers.

\textbf{FPGA platform.}
The accelerator is written in SystemVerilog RTL and implemented using AMD/Xilinx's Vitis design flow targeting 300\,MHz on a Xilinx Alveo U50 with HBM2; after place-and-route, the final design closes timing at 287\,MHz. The datapath parallelism parameters (Section~\ref{sec:accel-arch}) yield 2,048 INT8 MACs/cycle peak throughput. The deployment pipeline (Section~\ref{sec:deployment-pipe}) quantizes the model to INT8 using per-layer calibration over 300 frames, packs data into tiled layouts, and generates the NISA instruction stream with per-layer tile masks encoded as 128-bit bitmasks.

\textbf{GPU baseline.}
Inference latency is measured on an NVIDIA RTX~4090\,D (1,321 INT8 TOPS) running the same model in PyTorch. 

\textbf{Evaluation protocol.} We structure evaluation around four experiments:  

\begin{figure}[t]
\centering
\includegraphics[width=0.70\columnwidth]{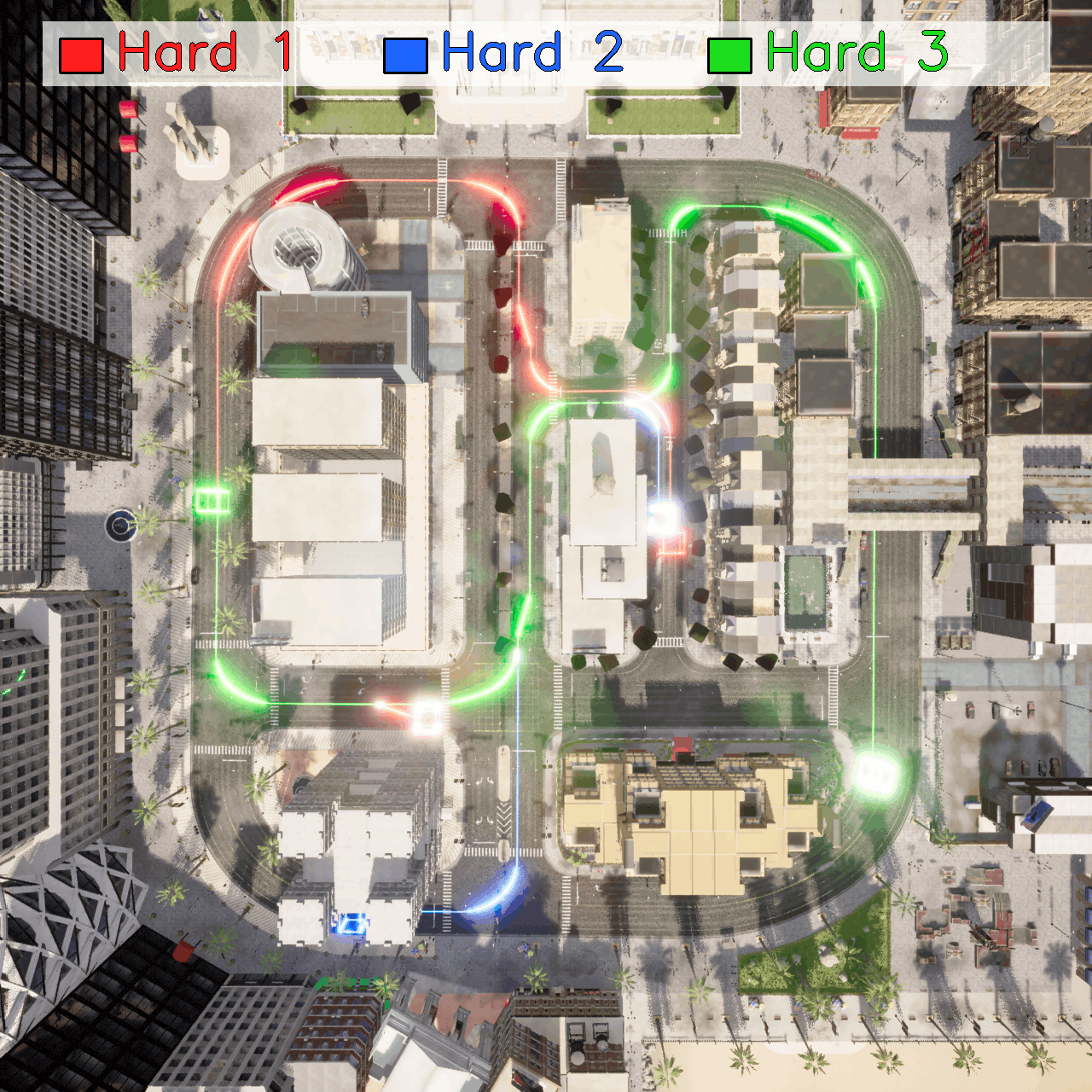}
\vspace{-0.1in}
\caption{Bird's-eye view of the three hard evaluation routes in \texttt{Town10HD\_Opt}. Squares mark spawn and destination points.}
\vspace{-0.2in}
\label{fig:eval_setup_bev}
\end{figure}

\textbf{E1. Driving quality — static 3-way comparison} (Section~\ref{sec:driving_quality}). We compare three configurations end-to-end in CARLA: (i)~dense baseline (GPU, float32), (ii)~task-sparse model (GPU, float32 with hard tile masks applied), and (iii)~task-sparse model deployed on-device on the Alveo U50 (INT8). All three run the same three hard routes under static conditions (0 NPC vehicles). Static evaluation eliminates the confound of CARLA's non-deterministic Traffic Manager, making a single run per configuration sufficient and exactly repeatable. \emph{Research question: does task-conditional sparsity preserve driving quality, and does INT8 on-device execution preserve the sparse model's behaviour?}

\textbf{E2. Driving quality — dynamic traffic robustness} (Section~\ref{sec:driving_quality}). We evaluate the task-sparse GPU model with 40 NPC vehicles over three independent runs and report mean and standard deviation. \emph{Research question: does the task-sparse model generalize to realistic dynamic traffic?}

\textbf{E3. Task-conditional vs.\ static pruning} (Section~\ref{sec:static_pruning}). We compare task-conditional masking against task-agnostic structured pruning at matched sparsity levels and equal fine-tuning budgets. \emph{Research question: is the quality benefit due to the conditional structure of the mask, or would any pruned model achieve the same result?}

\textbf{E4. FPGA on-device latency and efficiency} (Section~\ref{sec:latency}). Per-task latency and cycle decomposition are measured on the Alveo U50 via hardware performance counters embedded in the RTL, and compared against GPU inference to demonstrate that tile-level sparsity requires custom hardware to translate into actual speedup. \emph{Research question: what is the latency and energy benefit of task-conditional sparsity on the FPGA, and why does the same mask fail to help on a GPU?}

\textbf{Routes.}
All closed-loop experiments use three hard routes in CARLA \texttt{Town10HD\_Opt} shown in Figure~\ref{fig:eval_setup_bev}: Hard~1, a long multi-lane corridor with several intersections requiring lane changes and turns (206 waypoints); Hard~2, a dense commercial district with tight turns (129 waypoints); and Hard~3, a mixed residential--commercial path with multi-lane merges (269 waypoints).


\subsection{Task-Conditional Sparsity Analysis} \label{sec:task-cond-analysis}

The learned tile masks (Figure.~\ref{fig:hero} (a)) exhibit clear task-dependent structure. Table~\ref{tab:sparsity_latency} reports per-task sparsity and on-device FPGA performance for the final Phase~3 model.

Task-conditional sparsity reduces FLOPs by 66.3--76.1\% depending on the command. The masks reflect task complexity: Turn Left, which requires crossing oncoming traffic and navigating wider arcs, retains the most tiles (57/182). Brake, which relies primarily on proximity detection and requires minimal spatial processing, is pruned most aggressively (34/182). The remaining commands cluster tightly between 49 and 55 active tiles (26.9--30.2\%).

\textbf{Tile overlap.}
19 tiles (10.4\%) are active in \emph{all} tasks---a shared representational core. The union of all task masks covers 87 tiles (47.8\%), leaving 95 tiles (52.2\%) that are never activated by any command. The 68 task-differentiating tiles (active for some commands but not others) constitute the space where task-conditional sparsity provides its benefit over static pruning.

\rgen{This decomposes the savings into a static and a conditional part. Statically removing the 95 never-active tiles and keeping the union of the rest is a 52\% tile reduction, the same for every command and exactly what a task-agnostic structured prune achieves; task-conditional masking of the 68 then cuts a further 17--29\% depending on the command, reaching 69--81\% per command. To verify the two are separable, we removed the 95 never-active tiles from the architecture, producing a backbone with half the tiles (95 of 190), and retrained task-conditional masking on it from a freshly initialised gating network. Removing the 95 is lossless by construction---the full model under the union mask is bit-identical to the pruned model---and after retraining the smaller backbone the gating re-learns task-conditional masks with the same coarse ordering (braking pruned most, turns and lane changes least)---not the identical tiles, since a freshly initialised gater on the restructured backbone finds its own, sparser per-command solution, but the same qualitative structure---at preserved fidelity: open-loop control error 0.0267 matches the dense model, and the pruned model completes Hard~2 and Hard~3 at 100\%. The static removal and the task-conditional masking are therefore separable and additive: the 95 tiles are pure static redundancy, and the per-command benefit lives entirely in the surviving tiles.}


\textbf{FLOP vs.\ tile sparsity.}
FLOP reduction exceeds tile reduction for most commands because the gating network prunes more aggressively in early high-FLOP layers (large spatial dimensions) than in later low-FLOP layers, producing FLOP-aware masks despite training with a uniform tile-count penalty.

\begin{figure}[t]
\centering
\includegraphics[width=0.95\columnwidth]{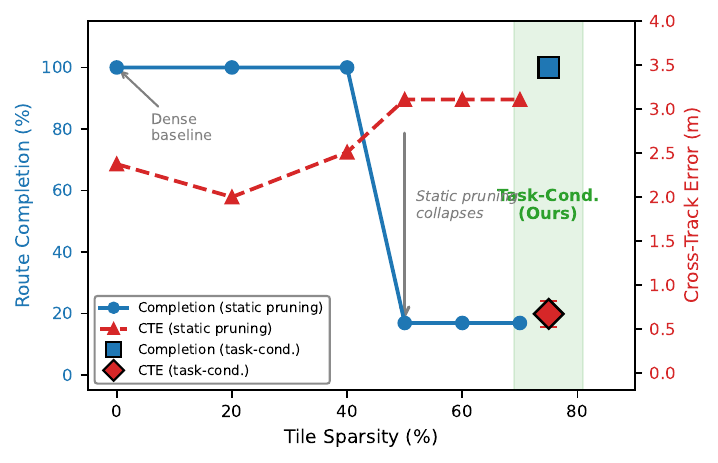}
\vspace{-0.15in}
\caption{Static pruning vs.\ task-conditional sparsity. Route completion (left, blue) and CTE (right, red) for static pruning at 0--70\% tile sparsity. Our task-conditional results (green band, 69--81\%) achieve 100\% completion with CTE 0.526--0.813 m, beyond the cliff where static pruning collapses}
\vspace{-0.15in}
\label{fig:static_pruning}
\end{figure}

\textbf{Latency breakdown.}
Compute cycles (Table~\ref{tab:sparsity_latency}) decrease with active tile count, confirming that masked tiles incur zero compute cost; the reduction is not strictly linear across commands because early-layer tiles (larger spatial maps) contribute disproportionately more compute per tile than later-layer tiles. Stall cycles, by contrast, decrease by only 24--26\% across all sparse tasks relative to the dense baseline. This asymmetry arises because stall is dominated by fixed per-layer costs---instruction fetch, DMA setup, and HBM access latency---that are independent of tile count. The dominant stall component is the pipeline barrier (compute idle while the next tile's DMA completes), which accounts for 84.7\% of stall in the dense baseline. Task-conditional sparsity reduces end-to-end latency by 51--59\%, from 9.12\,ms to 3.74--4.44\,ms.

\subsection{Static Pruning Comparison}\label{sec:static_pruning}
To isolate the benefit of task-conditional masks over task-agnostic pruning, we compare against static structured pruning at the same sparsity levels with equal fine-tuning budgets. For each target sparsity, we rank all 182 prunable tiles by weight L1 magnitude across the dense baseline, zero out the weakest tiles, and fine-tune for 20 epochs with pruned tiles held at zero. 

The results in Figure.~\ref{fig:static_pruning} show that static pruning maintains full route completion up to 40\% tile sparsity but collapses at 50\% (91 tiles), dropping to 16.9\% completion and remaining there through 70\% sparsity. The 16.9\% completion corresponds to the vehicle reaching the first required turn on each evaluation route and failing to execute it: at 50\%+ uniform sparsity, the statically pruned model loses turn-specific feature tiles and can only drive straight, terminating when CTE exceeds the 20m threshold. Our task-conditional approach operates at 69--81\% sparsity (34--57 active tiles per command), well beyond the static pruning cliff, while achieving 100\% completion on Hard~2 and Hard~3 with CTE of 0.526--0.813\,m (static evaluation, E1).

The failure mode is instructive. Static pruning must retain tiles that are important for \emph{any} command, while tiles that are important for \emph{some} commands but not others face an irresolvable trade-off. At 50\% sparsity the fixed mask inevitably removes tiles critical for specific commands (e.g., spatial-processing tiles needed for turns), collapsing performance on those tasks. Task-conditional masking avoids this by activating different tile subsets per command: the 68 task-differentiating tiles are retained when needed and skipped when not, enabling higher effective sparsity without sacrificing any individual command.

\begin{table}[t]
\centering
\caption{FPGA resource utilization (Alveo U50, 287\,MHz).}
\vspace{-0.1in}
\label{tab:utilization}
\begin{tabular}{lrrr}
\toprule
\textbf{Resource} & \textbf{Used} & \textbf{Available} & \textbf{Util.\%} \\
\midrule
CLB LUTs      & 235,048 & 870,720   & 27.0\% \\
CLB Registers & 389,479 & 1,743,360 & 22.3\% \\
BRAM36        & 238.5   & 1,344     & 17.8\% \\
URAM288       & 60      & 640       &  9.4\% \\
DSP48E2       & 2,082   & 5,952     & 35.0\% \\
\bottomrule
\end{tabular}
\vspace{-0.1in}
\end{table}

The 19-tile shared core (Section~\ref{sec:task-cond-analysis}) is structurally analogous to what static pruning would retain at high sparsity, but 19 tiles alone cannot drive the vehicle. The remaining capacity must come from task-specific tiles, which only the conditional mechanism can selectively enable.

\subsection{On-Device Latency}\label{sec:latency}
Table~\ref{tab:sparsity_latency} shows the per-task latency and cycle decomposition. Task-conditional sparsity reduces latency by 51--59\%, from 9.12\,ms (dense) to 3.74--4.44\,ms. Turn Right (55 prunable tiles) is the slowest sparse task at 4.44\,ms despite having fewer tiles than Turn Left, because it activates more tiles in high-FLOP early layers (larger spatial feature maps); Brake (34 tiles) is fastest at 3.74\,ms.

\subsection{GPU Comparison}
On the RTX~4090\,D, dense inference takes 0.44\,ms. Sparse inference with tile masking takes 0.54\,ms---22\% \emph{slower}. The mask multiplies activations by zero \emph{after} the convolution is already computed; the GPU still fetches all weights and occupies all CUDA cores. \rgen{Naive masking is not the only option, so we measured the best case for a custom kernel that truly skips the masked tiles: for each command we built a smaller dense network keeping only the active channels for that command (no masking, no gather, contiguous weights), which upper-bounds any sparse kernel, and timed it at batch one. Even after removing $\sim$75\% of the convolution , it runs within a few percent of the dense model---a mean 1.03$\times$ speedup, against the 2.1--2.4$\times$ the FPGA gets from the same masks. The limit is architectural: at batch one the network runs at well under 1\% of the GPU's peak, so latency is set by kernel launch and memory traffic across the 13 layers rather than arithmetic, and the pruned tiles sit mostly in deep, low-spatial layers that are already cheap on a GPU.} On the FPGA, masked tiles are skipped \emph{before} any DMA or compute is initiated, and its latency scales with the number of tile iterations, so the same mask that adds pure overhead on the GPU yields proportional cycle savings: tile-level sparsity needs hardware that drops fetch and compute together.

The 21$\times$ absolute latency gap (0.44 vs.\ 9.12\,ms) reflects the throughput difference (1,321 vs.\ $\sim$1.2 INT8 TOPS). The FPGA targets deployments where a discrete GPU is unavailable or where power is constrained, as analyzed next.

\subsection{Resource Utilization and Power}

All resource types (Table~\ref{tab:utilization}) are at or below 35\%, leaving headroom for scaling parallelism. The tile-level sparsity mechanism adds no additional datapath hardware; the same design executes both dense and sparse workloads (controlled through \texttt{sparse\_en}).

\textbf{Power.}
Vivado estimates 28.8\,W total on-chip power (22.6\,W dynamic, 6.1\,W static), well within the U50's 75\,W board budget. The kernel dynamic power (excluding the platform shell) is 17.7\,W. At 9.12\,ms per dense inference, energy is \rgen{263\,mJ}. Task-conditional sparsity reduces this to \rgen{108--128\,mJ} (51--59\% savings), driven primarily by faster completion rather than reduced instantaneous power. For comparison, the RTX~4090\,D at 425\,W TDP consumes up to $\sim$\rfix{187\,mJ} per inference---46--73\% more than the FPGA under sparse execution, despite being 21$\times$ faster in wall-clock time.


\begin{figure}[t]
\centering
\includegraphics[width=0.95\columnwidth]{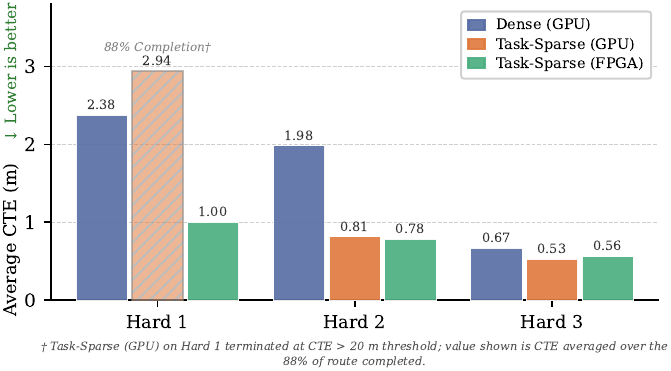}
\vspace{-0.1in}
\caption{Average CTE across three hard routes (E1: static, 0 NPC vehicles, single deterministic run). Task-Sparse (FPGA): INT8 on-device, verified against GPU float32 (max deviation 0.0129). Hatched bar: Task-Sparse (GPU) terminated early on Hard~1 (87.6\% completion, CTE averaged to termination). Task-Sparse models reduce CTE by 21--59\% vs.\ dense.}
\vspace{-0.2in}
\label{fig:driving_quality}
\end{figure}

\subsection{Driving Quality}\label{sec:driving_quality}

\textbf{Evaluation design.}
We evaluate under static conditions (E1: 0 NPC vehicles) to isolate the effect of sparsity and INT8 quantisation from traffic stochasticity, and complement this with dynamic-traffic results for the sparse GPU model (E2) to confirm robustness under realistic conditions.

\textbf{FPGA deployment fidelity.}
We verified on-device INT8 output against the GPU float32 model over 300 calibration frames; the maximum absolute difference across all six branch heads is 0.0129 (on a $[-1,+1]$ scale), confirming negligible quantization error.

\textbf{Results.}
Figure~\ref{fig:driving_quality} reports route completion and CTE across all three hard routes. On Hard~2 and Hard~3, all models achieve 100\% completion. The task-sparse models achieve substantially lower CTE than the dense baseline (59\% lower on Hard~2; 21\% lower on Hard~3). The FPGA on-device model matches the sparse GPU model within 3.9\% on Hard~2 and 5.9\% on Hard~3. On Hard~1, the longest route with several complex merges, the dense and FPGA models complete the route, while the sparse GPU model terminates early when lateral error exceeds 20\,m (87.6\% completion), indicating a localised failure at a specific merge geometry that the INT8 quantization noise incidentally avoids. \rgen{This is not a property of the task-conditional masking: the same mask, executed in INT8 on the FPGA, completes the route, so the early termination reflects a localised FP32-versus-INT8 numerical difference at one merge geometry, not the sparsity mechanism or data imbalance.}

\begin{figure}[t]
\centering
\includegraphics[width=\columnwidth]{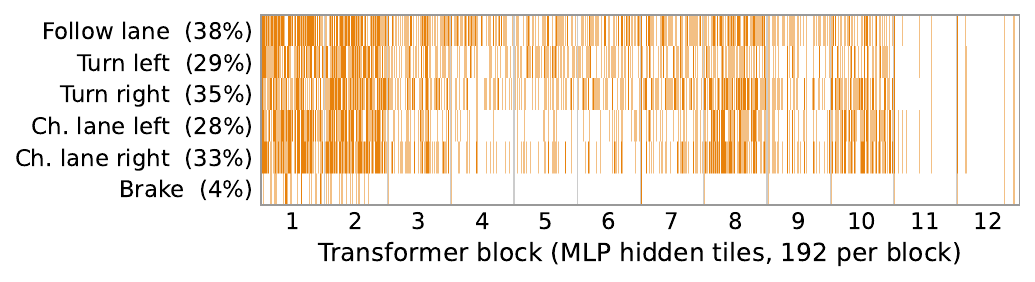}
\vspace{-0.25in}
\caption{\rgen{Learned per-command tile activation on the ViT-Base backbone (orange $=$ active; one column per MLP-hidden tile, grouped by transformer block). The gating network reproduces the CNN's structure: braking is pruned most, turns and lane-following least, and deeper blocks are pruned more.}}
\label{fig:vit_heatmap}
\vspace{-0.1in}
\end{figure}

\textbf{Effect of task-conditional masking.}
The CTE improvement of the sparse model is consistent with a task-specialization effect~\cite{shazeer2017moe}: masking irrelevant tiles reduces cross-command interference, sharpening per-command trajectory output. \rgen{The sparse model also undergoes additional Phase~2--3 fine-tuning, so to separate the conditional-architecture benefit from continued training we trained the dense baseline for the same extra budget and evaluated both over five runs on the three hard routes. The equal-budget dense model completes 75\%\,$\pm$\,13\% of routes (CTE 1.70\,$\pm$\,0.74\,m), whereas the task-sparse model completes 98\%\,$\pm$\,2\% (CTE 0.96\,$\pm$\,0.18\,m): continued training of a unified backbone does not reach the sparse model's driving quality and is markedly less stable run-to-run. The improvement is therefore attributable to conditional masking removing cross-command interference, not to the additional training.}

\begin{table*}[t]
\centering
\caption{Sparsity mechanism comparison. Our approach is the only design that conditions sparsity on a runtime task signal with zero datapath modification. $^\dagger$Compressed format with index matching.}
\vspace{-0.1in}
\label{tab:sparsity_comparison}
\resizebox{\textwidth}{!}{
\begin{tabular}{lccccccc}
\toprule
\textbf{Design} & \textbf{Venue} & \textbf{Sparsity Type} & \textbf{Granularity} & \textbf{Pattern Source} & \textbf{Runtime Adaptive?} & \textbf{HW Overhead} & \textbf{Datapath Change?} \\
\midrule
Eyeriss~\cite{chen2016eyeriss}       & ISCA'16  & Activation zeros    & Element     & Input data  & Per-input & Gating logic       & No  \\
SCNN~\cite{parashar2017scnn}         & ISCA'17  & Weight + activation & Element     & Training    & No        & Index match$^\dagger$ & Yes \\
Cambricon-S~\cite{zhou2018cambricon} & MICRO'18 & Weight (coarse)     & Filter      & Training    & No        & Index decoder      & Yes \\
SNAP~\cite{snap2021}              & JSSC'21 & Weight + activation      & Element    & Training    & No        & Bitmap index match$^\dagger$ & Yes \\
HPIPE~\cite{hall2020hpipe}           & FPGA'20  & Weight sparsity     & Element     & Pruning     & No        & Skip logic         & Yes \\
\midrule
\textbf{Ours}                        & ---      & \textbf{Task-cond.\ tile} & \textbf{16-ch tile} & \textbf{Task command} & \textbf{Per-task} & \textbf{1-bit check} & \textbf{No} \\
\bottomrule
\end{tabular}
}
\vspace{-0.1in}
\end{table*}

\textbf{Dynamic traffic.}\label{sec:dynamic_eval}
Under dynamic conditions (E2: 40 NPC vehicles, three independent runs), the task-sparse GPU model achieves 100\% completion on Hard~2 and Hard~3 in all runs, with CTE of $0.710 \pm 0.014$\,m and $0.520 \pm 0.014$\,m respectively---consistent with static results and confirming robustness to moderate traffic density. Hard~1 exhibits run-to-run completion variance (61.7--65.0\%) attributable to the combination of the model's Hard~1 instability and stochastic traffic divergence over the long route.

\subsection{\rgen{Generality to Transformer Backbones}}\label{sec:generality_vit}
\rgen{To test whether the mechanism generalises beyond convolutions, we built a second visuomotor controller on the same CARLA task with a ViT-Base/16 backbone in place of the CNN, keeping the gating network, the three-phase training, and the tile-level sparsity penalty unchanged. The masking target is the MLP, the high-FLOP part of a transformer block (\texttt{fc1} and \texttt{fc2} are 64\% of a block's FLOPs, attention the other 36\%): we place the tiles on the MLP hidden channels---the output channels of \texttt{fc1}, which are the input channels of \texttt{fc2}---so masking a tile removes work from both linear layers, the same output-side and input-side skip the CNN gets across adjacent convolutional layers.}

\rgen{Figure~\ref{fig:vit_heatmap} shows the learned masks. The gating network reproduces the CNN's ordering: braking, the simplest command, is pruned most (3.6\% of tiles active), while turns and lane-following keep the most (28--38\%), with $\sim$28\% active overall, and deeper blocks pruned more heavily as in the CNN. Open-loop fidelity matches the dense ViT: the mean absolute error over predicted steering and acceleration is 0.030 versus 0.028. Because we mask only the MLP and leave attention dense, the per-command FLOP reduction is 40--62\% (Table~\ref{tab:vit}), below the CNN's 66--76\%; attention is a floor we do not touch, and masking attention head-groups (standard head pruning~\cite{michel2019sixteen}) would lift this further. The result needs no new accelerator: the engine already executes GEMM---the linear layers are 96\% of a ViT block's FLOPs (the MLP alone is 64\%)---and a GEMM is a 1$\times$1 convolution on this engine, so the same instruction-level masks apply.}


\begin{table}[t]
\centering
\caption{\rgen{Per-command sparsity and FLOP reduction on the ViT-Base backbone. Active tiles are of the MLP hidden dimension; attention is left dense.}}
\label{tab:vit}
\vspace{-0.1in}
\small
\begin{tabular}{lcc}
\toprule
\textbf{Command} & \textbf{Active tiles} & \textbf{FLOP reduction} \\
\midrule
Follow lane     & 37.9\% & 39.7\% \\
Turn left       & 29.0\% & 45.4\% \\
Turn right      & 34.9\% & 41.6\% \\
Ch.\ lane left  & 28.2\% & 45.9\% \\
Ch.\ lane right & 33.2\% & 42.7\% \\
Brake           & \phantom{0}3.6\% & 61.6\% \\
\bottomrule
\end{tabular}
\vspace{-0.15in}
\end{table}

\subsection{Comparison with Prior Work}
\textbf{Sparsity mechanism comparison.}
Table~\ref{tab:sparsity_comparison} compares our sparsity mechanism against prior accelerators along three dimensions: what determines the sparse pattern, at what granularity skipping occurs, and what hardware overhead the mechanism requires. Our approach is unique in conditioning the pattern on a task command with zero datapath overhead.

\textbf{FPGA accelerator comparison.}
Table~\ref{tab:fpga_comparison} compares our design against recent FPGA-based inference accelerators. The designs span different platforms, architectures, and compute paradigms; we report numbers directly from the respective publications.

\begin{table*}[t]
\centering
\caption{Comparison with FPGA-based inference accelerators. Our contribution is not peak throughput but the 2.1--2.4$\times$ latency reduction from task-conditional sparsity---a capability absent from all other designs.}
\vspace{-0.1in}
\label{tab:fpga_comparison}
\resizebox{\textwidth}{!}{
\begin{tabular}{llccccccl}
\toprule
\textbf{Design} & \textbf{Venue} & \textbf{Platform} & \textbf{Freq.} & \textbf{Dtype} & \textbf{DSPs} & \textbf{Peak (GOPS)} & \textbf{Open} & \textbf{Focus} \\
\midrule
Vitis AI DPU~\cite{xilinx2022dpu}       & --  & Alveo U280    & 300\,MHz  & INT8    & 2,452  & 4,800  & No  & Commercial DPU IP (multi-MAC/DSP packing) \\
DNNExplorer~\cite{zhang2021dnnexplorer}  & ICCAD'20     & Xilinx KU115  & 200\,MHz  & INT8/16 & 4,686  & 1,702  & No  & Hybrid pipeline+generic accelerator with automated DSE \\
HPIPE~\cite{hall2020hpipe}              & FPGA'20      & Stratix 10    & 580\,MHz  & INT16   & 5,022  & ---    & No  & Layer-pipelined sparse-aware CNN inference \\
H2PIPE~\cite{mario2024h2pipe}               & FPL'24       & Stratix 10 NX & 300\,MHz  & INT8    & 3,960  & 7,731  & No  & Layer-pipelined CNN with HBM weight offloading \\
CHARM~\cite{zhuang2023charm}            & FPGA'23      & VCK190        & 1\,GHz    & FP32    & AIE    & 2,936  & Yes & Heterogeneous AIE+PL GEMM composition \\
FINN-R~\cite{blott2018finnr}            & TRETS'18     & AWS F1 (VU9P) & 250\,MHz  & INT1--3 & ---$^\dagger$ & ---  & Yes & Streaming dataflow for quantised NNs \\
Sparse DLA~\cite{sparsedla2021}         & FCCM'21      & Stratix 10 MX & 257\,MHz  & FP16    & 1,442  & ---    & No  & Sparse matrix packing for systolic array with HBM2 \\
HASS~\cite{hass2024}                    & FPL'24       & Alveo U250    & 250\,MHz  & INT16   & 7,434  & ---    & No & HW-aware sparsity search for dataflow accelerators \\

\midrule
\textbf{Ours}                          & ---             & \textbf{Alveo U50}   & \textbf{287\,MHz}  & \textbf{INT8}  & \textbf{2,082} & \textbf{1,176} & \textbf{Yes} & \textbf{Task-conditional tile sparsity} \\
\bottomrule
\multicolumn{9}{l}{\small $^\dagger$FINN-R uses LUT-based compute rather than DSPs for sub-INT8 quantisation.} \\
\multicolumn{9}{l}{\small Ours: 9.12\,ms dense $\rightarrow$ 3.74--4.44\,ms sparse (2.1--2.4$\times$ speedup); \rgen{263\,mJ $\rightarrow$ 108--128\,mJ} energy/inference; 110 $\rightarrow$ 226--267\,inf/s.} \\
\end{tabular}
}
\vspace{-0.2in}
\end{table*}

Several observations contextualize these numbers. The AMD DPU achieves higher peak GOPS by packing multiple INT8 MACs per DSP48E2~\cite{xilinx2022dpu}; our design uses one MAC per DSP, a conservative choice upgradeable orthogonally to tile masking. DNNExplorer generates per-model customized accelerators with 87--90\% DSP efficiency on the larger U250 FPGA~\cite{zhang2021dnnexplorer}. HPIPE targets Intel Stratix~10 FPGAs with INT16 precision and supports static weight sparsity through channel-level skipping~\cite{hall2020hpipe}. CHARM leverages Versal AI Engine cores at 1.25\,GHz---a fundamentally different compute fabric than programmable logic DSPs~\cite{zhuang2023charm}. FINN-R generates fully-pipelined streaming architectures using LUTs rather than DSPs, optimized for sub-INT8 quantization and achieving high throughput at extremely low power~\cite{blott2018finnr}.

Our design occupies a distinct point in this space: a general-purpose tiled accelerator with \emph{runtime-adaptive} sparsity controlled through the instruction stream, deployed on a mid-range FPGA. No other design supports task-conditional execution masks. The closest is HPIPE's static weight sparsity~\cite{hall2020hpipe}, which is fixed at compile time and cannot adapt per task. On our hardware, task-conditional masking reduces latency by 51--59\% (9.12\,ms $\rightarrow$ 3.74--4.44\,ms), yielding 226--267 inferences per second. At 28.8\,W board power, sparse inference consumes \rgen{108--128\,mJ}, a 51--59\% energy reduction from the dense baseline. A GPU runs the same model in 0.44\,ms dense, but the sparse variant is 22\% \emph{slower} (Section~4.5), confirming that tile-level sparsity requires hardware support. Please note that our throughput can be improved by design scaling, DSP-packing, and various other optimizations orthogonal to our work.

\section{Related Work}

\textbf{Sparse accelerators.}
Prior work exploits weight and activation sparsity through compressed encodings and index-matching hardware (SCNN~\cite{parashar2017scnn}, SparTen~\cite{gondimalla2019sparten}), coarse-grained filter pruning with dedicated index decoders (Cambricon-S~\cite{zhou2018cambricon}), flexible dataflows adapting to varying sparsity patterns (Eyeriss v2~\cite{chen2019eyeriss}), unstructured sparsity (SNAP~\cite{snap2021}), and irregular sparsity mapping via flexible interconnects (SpArch~\cite{zhang2020sparch}). \rgen{A complementary line schedules tile-level sparsity across heterogeneous cores rather than encoding it: DeSpa~\cite{jang2025despa} pairs dense and sparse cores with a tile-stealing scheduler that adapts to the sparsity already present in the data. Such heterogeneity-aware scheduling accelerates sparse tiles but cannot skip task-irrelevant ones, because it has no task signal and sees only the tensor; our command-conditioned masks instead tell the hardware, before any fetch, which tiles not to execute at all, and the two are composable---one could schedule the surviving unmasked tiles across DeSpa-style cores.}

All of these designs treat sparsity as a \emph{static} property determined at training or pruning time. The sparse pattern does not change based on what the model is being asked to do at runtime. Our work differs in that the sparsity pattern is \emph{task-conditional}: different tasks activate different tiles, and the mask is resolved at no runtime cost. 

\textbf{Dynamic and conditional computation.}
A parallel line of work explores input-dependent computation on standard processors. SkipNet~\cite{wang2018skipnet} and BlockDrop~\cite{wu2018blockdrop} learn per-input policies that skip entire residual blocks, achieving FLOP reductions of 30--40\% on ImageNet. FBS~\cite{gao2019fbs} dynamically prunes channels based on intermediate activations, and early-exit networks~\cite{teerapittayanon2016branchynet} terminate inference at an intermediate layer when confidence is high. Mixture-of-Experts (MoE) architectures~\cite{shazeer2017moe, fedus2022switch} route tokens to specialized subnetworks, activating only a subset of parameters per input.

These approaches demonstrate that computation can be safely skipped based on runtime signals, but they are designed for GPU/CPU execution where ``skipping'' means zeroing outputs or branching around computation---the hardware still fetches weights, allocates registers, and occupies memory bandwidth. Without alignment to the accelerator's tiling granularity, the FLOP reduction does not translate to proportional speedup. Our gating mechanism differs in two respects: (1) it conditions on the \emph{task command} rather than on input features, avoiding any per-input overhead; and (2) it produces masks at the exact granularity of the hardware's tile iteration, enabling zero-cost skipping on the accelerator.

\textbf{Structured pruning.}
Static structured pruning removes entire filters~\cite{li2017pruning}, channels, or blocks at training time to reduce model size and compute~\cite{han2015learning, han2016deep, wen2016learning}. These techniques produce a single pruned model that is deployed uniformly for all inputs. Our approach can be viewed as a form of \emph{conditional} structured pruning, where the pruning pattern varies per task. The 68 task-differentiating tiles reveal that a single static pruning pattern cannot capture the task-dependent structure that the gating network learns.



\textbf{End-to-end autonomous driving.}
Conditional imitation learning~\cite{codevilla2018end} introduced the branched architecture we build upon, where a high-level command selects among specialized output heads sharing a common backbone. Subsequent work has explored richer command representations~\cite{codevilla2019exploring}, attention mechanisms, and transformer-based architectures for driving. The DAgger algorithm~\cite{ross2011dagger} addresses covariate shift in imitation learning by iteratively collecting data under the learned policy's state distribution. CARLA~\cite{dosovitskiy2017carla} provides the simulation environment for training and evaluating driving controllers in diverse traffic scenarios. Our work applies these driving models as an evaluation vehicle for the co-designed sparsity mechanism, demonstrating that task-conditional sparsity is not just a theoretical construct but produces measurable latency and energy savings on real hardware executing a real control task.

\section{Conclusion}

We presented a HW/SW co-designed system that exploits the task command in multi-task inference models to achieve structured sparsity with zero runtime overhead. A lightweight gating network ($<$0.12\% of backbone parameters) maps a one-hot task vector to per-tile binary execution masks aligned with the accelerator's scheduling granularity. The masks are encoded into the instruction stream as bitmask fields, enabling the tile manager to skip masked tiles entirely---no weight fetch, no activation load, no compute---at the cost of a single bit check per tile.

We prototyped the system on a Xilinx Alveo U50 FPGA with HBM, deploying an 8-layer CNN backbone with six branched heads for closed-loop visuomotor driving in the CARLA simulator. Task-conditional sparsity reduces FLOPs by 66--76\% and on-device latency by 51--59\% across six driving commands, while the task-sparse model maintains 100\% route completion. The learned masks reflect task semantics: left turns retain the most computation; braking is pruned most aggressively. On the same model, a GPU executes the sparse variant 22\% \emph{slower} than dense, validating that tile-level sparsity requires hardware support to yield actual speedup.

The approach generalizes beyond the specific model and task evaluated here. Command-conditioned multi-task models with a discrete task selector---robotic manipulation with grasp-type commands, multi-language speech recognition, or multi-objective optimization --- can leverage the same mechanism. \rgen{The ISA's tile mask fields are opcode-agnostic and apply to convolutions and the linear layers of transformers, which we confirm on a ViT-Base backbone.}

Future work includes extending the gating mechanism to support \emph{input-conditional} sparsity~\cite{verelst2020dynamic} (where the mask depends on both the task and the input image), exploring hierarchical masking at multiple granularities (block, layer, tile, channel), and scaling the accelerator to larger models and higher parallelism configurations on multi-SLR FPGA platforms.

\bibliographystyle{ACM-Reference-Format}
\bibliography{main}

%
%
%
%
%

\appendix
\section{Artifact Appendix}

\sloppy
\emergencystretch=3em

\subsection{Abstract}

This artifact reproduces the two on-device results tables of \emph{Sparse by
Command}: \textbf{Table~2} (per-command active-tile counts and FPGA inference
latency/speedup) and \textbf{Table~3} (accelerator resource utilization and
clock frequency). It provides the accelerator RTL, the C++/OpenCL host sources,
the pre-built bitstream (287~MHz, INT8), the quantized INT8 weights/biases and
golden references, the per-command tile masks, the trained dense and sparse
model checkpoints, and the CARLA training/evaluation scripts. The gating network
maps each one-hot driving command to per-tile execution masks; one script
reproduces the active-tile counts (49/57/55/54/53/34 of the 182 prunable tiles)
exactly, a second builds the host and runs it against the bitstream on an
Alveo~U50 to reproduce the per-command latency and speedup, and a third reports
the Table~3 utilization from the shipped build logs (or, optionally, from a fresh
Vivado/Vitis build). The paper's central claim---task-conditional compute
skipping via per-command tile masks---is reproduced quantitatively by Table~2.

\subsection{Artifact check-list (meta-information)}

{\small
\begin{itemize}
  \item {\bf Algorithm: } Task-conditional structured (tile-level) sparsity for tiled inference accelerators.
  \item {\bf Program: } SystemVerilog RTL accelerator; C++/OpenCL~1.2 host; Python scripts (active-tile extraction; CARLA training/eval).
  \item {\bf Compilation: } Vivado 2022.1 (package \texttt{.xo}); Vitis \texttt{v++} (link \texttt{.xclbin}); \texttt{g++} (C++14) with XRT/OpenCL.
  \item {\bf Transformations: } None at run time (masks precomputed and encoded into the instruction stream).
  \item {\bf Binary: } Pre-built \nolinkurl{vla_accel.xclbin} (Alveo~U50, 287~MHz). The host is \emph{built from source} on the target machine (it links against the local XRT), so no host binary is shipped.
  \item {\bf Model: } 8-layer INT8 CNN driving controller with six branched heads and a lightweight tile-gating MLP. Both the dense and the sparse (gated) checkpoints are included.
  \item {\bf Data set: } The quantized INT8 weights, INT32 biases, and per-layer golden reference outputs (used to verify INT8 correctness); see \S\ref{sec:ae-data}.
  \item {\bf Run-time environment: } Ubuntu with XRT 2.14 (2022.2).
  \item {\bf Hardware: } AMD/Xilinx Alveo~U50.
  \item {\bf Run-time state: } The card is reset to a known HBM-calibration state before latency measurement (see \S\ref{sec:ae-eval}).
  \item {\bf Execution: } Scripted; one inference pass per configuration; on-chip performance counters read back over PCIe.
  \item {\bf Metrics: } Per-command active-tile counts; on-device latency and total/compute cycles; speedup; INT8 correctness; and DSP, BRAM, URAM, LUT, and FF utilization with achieved clock frequency.
  \item {\bf Output: } A console table of per-command active tiles, cycles, latency, and speedup with pass/fail INT8 verification (Table~2), and the parsed utilization report (Table~3).
  \item {\bf Experiments: } E1 active-tile counts (software); E2 on-device latency (FPGA); E3 resource utilization and frequency (build logs / rebuild).
  \item {\bf How much disk space required (approximately)?: } $\sim$2\,GB with the pre-built bitstream; $\sim$10\,GB if rebuilding from RTL.
  \item {\bf How much time is needed to prepare workflow (approximately)?: } $\sim$15--30\,min with the pre-built bitstream.
  \item {\bf How much time is needed to complete experiments (approximately)?: } $\sim$10\,min (E1+E2); $\sim$3\,h for an optional bitstream rebuild (E3).
  \item {\bf Publicly available?: } Yes.
  \item {\bf Code licenses (if publicly available)?: } MIT License.
  \item {\bf Data licenses (if publicly available)?: } The model checkpoints, quantized weights, and golden references are released under the same license as the code; no third-party dataset is redistributed.
  \item {\bf Workflow automation framework used?: } Shell + Python scripts (no external framework).
  \item {\bf Archived (provide DOI)?: } https://doi.org/10.5281/zenodo.21503283
\end{itemize}
}

\subsection{Description}

\subsubsection{How to access}
The artifact is located at \url{https://doi.org/10.5281/zenodo.21503283}, \nolinkurl{sparse-by-command_artifacts/},
containing: the accelerator RTL (\nolinkurl{rtl/}), the Vivado/Vitis build scripts
(\nolinkurl{hw_build/}), the C++/OpenCL host sources and per-command specs
(\nolinkurl{host/}), the software side---model definitions, dense and sparse
checkpoints, active-tile extractor, and CARLA training/eval scripts
(\nolinkurl{sw/}), one-shot reproduction scripts (\nolinkurl{scripts/}), and the
pre-built bitstream, quantized data, and the last full hardware-build reports
(\nolinkurl{prebuilt/}).

\subsubsection{Hardware dependencies}
An AMD/Xilinx Alveo~U50 accelerator card in a PCIe~Gen3~x16 host is required for
the on-device measurements (E2). The optional bitstream rebuild (E3) needs only a
machine that can run Vivado/Vitis; no FPGA is required for the build itself.

\subsubsection{Software dependencies}
For E2: Xilinx Runtime (XRT) 2.14 / 2022.2 and the U50 deployment shell, with
\texttt{g++} (C++14) and OpenCL~1.2 headers to build the host. For the optional
E3 rebuild: Vivado~2022.1 and Vitis~\texttt{v++} (the design also builds under
2024.2, at a slightly lower achieved frequency). For the software side: Python3
with PyTorch (active-tile extraction) and, for the provided training/evaluation
recipes, the CARLA~0.9.14 simulator and its Python API together with their usual
dependencies (\texttt{numpy}, \texttt{opencv}, \texttt{pygame}).

\subsubsection{Data sets}
\label{sec:ae-data}
The artifact provides pre-trained checkpoints and the derived tensors: the per-layer quantized INT8 weights, INT32 biases, and golden reference activations (\nolinkurl{fpga_data_carla/}), extracted from the
CARLA \nolinkurl{Town10HD_Opt} calibration frames. The $\sim$88 golden tensors let the host verify INT8 correctness bit-exactly during E2.

\subsubsection{Models}
The 8-layer CNN backbone with six branched heads and the tile-gating MLP; model definitions are in \nolinkurl{sw/models/}. Both trained checkpoints are included in \nolinkurl{sw/checkpoints/}: the dense controller and the sparse (gated) controller whose gater produces the per-command tile masks. The corresponding dense and sparse training and evaluation recipes are provided under \nolinkurl{sw/} for completeness.

\subsection{Installation}
No installation is needed beyond the target toolchains. Two scripts drive the
FPGA flow. \texttt{scripts/reproduce\_table2.sh} \emph{builds the host} from
source (\nolinkurl{host/make_host.py} compiles \nolinkurl{main_carla.cpp} into
\nolinkurl{app_single.exe}) and then runs that host against
\nolinkurl{vla_accel.xclbin}, streaming the quantized weights/biases and the
per-command tile masks to the card and reading back the on-chip performance
counters. \texttt{scripts/build\_bitstream.sh} (optional) regenerates the
bitstream: it packages the RTL into \nolinkurl{vla_accel.xo} and links it into
\nolinkurl{vla_accel.xclbin}, emitting the Table~3 reports.

\subsection{Experiment workflow}
\begin{itemize}
  \item \textbf{E1 --- active-tile counts (Table~2, software).}\\
        \texttt{cd sw \&\& python3 extract\_active\_tiles.py}\\
        loads the sparse gating checkpoint, runs the gating MLP on each one-hot
        command, thresholds at 0.5, and prints the active-tile count per command.
  \item \textbf{E2 --- on-device latency (Table~2, FPGA).}\\
        \texttt{DO\_RESET=0 bash scripts/reproduce\_table2.sh}\\
        builds the host, resets the card to the paper HBM-calibration state, then
        runs the dense configuration and each of the six commands, printing active
        tiles, total/compute cycles, latency, speedup, and INT8 verification.
  \item \textbf{E3 --- resource utilization and frequency (Table~3).} Two methods:
        \emph{(a) prebuilt (default).}
        \texttt{bash scripts/reproduce\_table3\_prebuilt.sh}
        parses the shipped placement report and prints the DSP, BRAM, URAM, LUT,
        and FF counts. \emph{(b) rebuild (optional, $\sim$3\,h).}
        \texttt{bash scripts/build\_bitstream.sh}
        re-runs the full build and points to the freshly generated reports.
\end{itemize}

\subsection{Evaluation and expected results}
\label{sec:ae-eval}
\textbf{E1} prints \texttt{49/57/55/54/53/34} (of 182), matching the Table~2
``Active Tiles'' column exactly. \textbf{E2} prints, per command, the active
tiles, total/compute cycles, latency, and speedup: the dense configuration is
$\approx$2.62\,M cycles ($\approx$9.12\,ms at 287\,MHz), the per-command speedups
fall in the 2.1--2.4$\times$ range of Table~2, and every run reports
\texttt{TEST PASSED} against the golden references. \textbf{E3} reproduces
Table~3: DSP~2082, BRAM~238.5, and URAM~60 exactly, with LUT and FF within
$<$0.3\% of the reported values, at an achieved clock of $\approx$287\,MHz. Method
(a) reads these from the shipped report at
\nolinkurl{prebuilt/hw_reports/utilization_placed.rpt}; method (b) reads them from
the \nolinkurl{impl_1} report emitted by the rebuild.

\paragraph{Reproducibility note (FPGA latency).}
Compute cycles, active-tile counts, speedup ratios, INT8 correctness, and the
Table~3 utilization are deterministic and reproduce exactly. The absolute total
cycles (hence latency) include an HBM memory-wait component whose value depends on
the HBM PHY timing selected at each card reset ($\sim\pm3\%$). The paper's Table~2
corresponds to the ``slow'' calibration (dense $\approx$2.62\,M cycles
$\rightarrow$ 9.12\,ms); \texttt{reproduce\_table2.sh} resets the card until that
state is reached and then reproduces the reported latencies within $\sim$1--2\%.

\paragraph{Scope (closed-loop driving, Figures 8 and 9).}
The artifact includes the CARLA training and evaluation scripts, but the
closed-loop \emph{driving} results (Figures~8 and~9) are \emph{not} claimed as
reproducible artifacts. CARLA's per-run rendering and physics are
non-deterministic and depend on the GPU model, the GPU load, and the host
workstation---a documented simulator behavior
(\url{https://github.com/carla-simulator/carla/issues/4004})---so absolute
cross-track error is not portably reproducible across machines. The paper's
core contribution---task-conditional compute skipping, i.e., cutting the active
compute per driving command via learned tile masks---is captured by the
deterministic on-device measurements of Table~2 (and its implementation cost by
Table~3), which are the focus of this artifact.

\subsection{Experiment customization}
The per-command masks can be swapped by pointing the host at a different
\nolinkurl{model_spec_task*.txt} spec to run any tile-sparsity pattern; setting
\texttt{DO\_RESET=0} makes \texttt{reproduce\_table2.sh} measure the current HBM
state without the reset loop. The parallelism parameters (\texttt{OC\_PAR},
\texttt{PP\_PAR}) are defined in \nolinkurl{host/srcs/globals.hpp} for rebuilds.

\subsection{Methodology}

Submission, reviewing and badging methodology:

\begin{itemize}
  \item \url{https://www.acm.org/publications/policies/artifact-review-and-badging-current}
  \item \url{https://cTuning.org/ae}
\end{itemize}

\end{document}